\documentclass[prd,preprint,amsmath]{revtex4}
\usepackage{bm} 
\usepackage{bm}
\newcommand{\beq}{\begin{equation}}
\newcommand{\eeq}{\end{equation}}
\newcommand{\beqa}{\begin{eqnarray}}
\newcommand{\eeqa}{\end{eqnarray}}
\newcommand{\beqar}{\begin{eqnarray*}}
\newcommand{\eeqar}{\end{eqnarray*}}

\newcommand{\hnabla}{\hat\nabla}

\newcommand\munu{\ensuremath{{\mu\nu}}}

\newcommand{\dspst}{\displaystyle}

\usepackage{bm}

\begin{document}

\title{A class of exact solutions of Einstein's field equations in
higher dimensional spacetimes, d${\bm\geq 4}$: Majumdar-Papapetrou
solutions}
\author{Jos\'e P. S. Lemos\thanks{e-mail:lemos@kelvin.ist.utl.pt}}
\affiliation{
Centro Multidisciplinar de Astrof\'{\i}sica --
CENTRA,
\\Departamento de F\'{\i}sica,
Instituto Superior T\'ecnico, 
Universidade T\'ecnica de Lisboa,
Av. Rovisco Pais 1, 1049-001 Lisboa, Portugal.}
\author{Vilson T. Zanchin\thanks{e-mail: zanchin@ccne.ufsm.br}}
\affiliation{Departamento de F\'{\i}sica, Universidade Federal de
Santa Maria, 97119-900 Santa Maria, RS, Brazil.}
\begin{abstract}
\noindent
The Newtonian theory of gravitation and electrostatics admit
equilibrium configurations of charged fluids where the charge density
can be equal to the mass density, in appropriate units. The general
relativistic analog for charged dust stars was discovered by Majumdar
and by Papapetrou.  In the present work we consider Einstein-Maxwell
solutions in d-dimensional spacetimes and show that there are
Majumdar-Papapetrou type solutions for all ${\rm d} \geq 4$.  It is verified
that the equilibrium is independent of the shape of the distribution
of the charged matter.  It is also showed that for perfect fluid
solutions satisfying the Majumdar-Papapetrou condition with a boundary
where the pressure is zero, the pressure vanishes everywhere, and that
the $({\rm d}-1)$-dimensional spatial section
of the spacetime is conformal to a Ricci-flat space.
The Weyl d-dimensional axisymmetric solutions are generalized to 
include electric field and charged matter.

\end{abstract}
\maketitle

%\newpage

%\centerline{}
%\newpage

\section{Introduction}
\label{introd}

Objects formed by elementary components with electric charge to mass
ratio equal to one have been considered for a long time now in general
relativity, and in other theories of gravity.  It is clear that the
Newtonian theory of gravitation and Coulomb electrostatics conjointly
admit equilibrium configurations for charged fluids where the
electrical charge density $\rho_{\rm e}$ is equal to the mass density
$\rho$, in appropriate units. Such a neutral equilibrium is possible
owing to the exact balancing of the gravitational and Coulombian
electric forces on every fluid particle.  Thus, a static distribution
of charged dust of any shape can in principle be built.  The general
relativistic analog for such extremal charged dust configurations was
discovered by Majumdar \cite{maj47} and independently by Papapetrou
\cite{papa47}.  First, Weyl \cite{weyl}, while studying the
electrostatic field in vacuum Einstein-Maxwell theory in an
axisymmetric static four-dimensional spacetime, found that if the
metric component $g_{tt}\equiv V(x^i)$ and the electric potential
$\phi(x^i)$ (where $x^i$ represent the spatial coordinates, $i=1,2,3$)
are related by a functional form $V=V(\phi)$, then this function is
given by
\beq
V= A + B\phi +\phi^2\, ,\label{weylcondition}
\eeq
where $A$ and $B$ are arbitrary constants, and we use geometrical 
units, $G=1$, $c=1$.  
Majumdar \cite{maj47}
extended this result by showing that it holds for a large class of
static spacetimes with no particular spatial symmetry, axial or
otherwise, for which the metric can be written as
\begin{equation}
ds^2 = - V dt^2 + h_{ij}\,dx^idx^j\, , 
\quad\quad i,j=1,2,3\;, \label{metric4d}
\end{equation}
where $V$ and $h_{ij}$ are functions of the spatial coordinates 
$x^i$.  Moreover, by choosing $B=\pm2\sqrt{A\,}$, in which case the
potential $V$ assumes the form of a perfect square,
\beq
V = \left(\sqrt{A\,}\pm\phi\right)^2\, \label{majumdar1}
\eeq
Majumdar was able to show that the Einstein-Maxwell equations in the
presence of charged dust imply exactly the same relation of the
Newtonian theory 
\beq 
\rho_{\rm e} = \pm \rho\, ,\label{majumdar2}
\eeq
with both the gravitational potential $V$ and the electric potential
$\phi$ satisfying a Poisson-like equation.  As in the Newtonian case,
the relativistic solutions are static configurations of charged dust
(a perfect fluid with zero pressure) and need not have any spatial
symmetry.  Majumdar \cite{maj47} also showed that in the case $V$ is a
perfect square as in Eq. (\ref{majumdar1}) the metric of the
three-space is conformal to a flat metric whose conformal factor is
given by $1/V$, and in such a case all the stresses in the charged
matter vanish.  Similar results were found by Papapetrou
\cite{papa47}, who assumed as starting point a perfect square relation
among $V$ and $\phi$.  Condition (\ref{majumdar1}) is called the
Majumdar-Papapetrou condition. We shall see that the condition
(\ref{majumdar1}) implies (\ref{majumdar2}), whereas the converse is
not generally true. Solutions in which conditions  (\ref{majumdar1}) 
and (\ref{majumdar2})
hold are called Majumdar-Papapetrou solutions. 

After the works of Majumdar \cite{maj47} 
and Papapetrou \cite{papa47}, several authors have
studied different aspects of static spacetimes satisfying the
Majumdar-Papapetrou conditions. In vacuum the extreme
Reissner-Nordstr\"om spacetime and the corresponding multi-black hole
solutions were analyzed first by Hartle and Hawking
\cite{hartlehawking} (see also \cite{azumakoikawa}). 
Other solutions in vacuum, are the
Israel-Wilson-Perjes rotating solutions \cite{israelwilson,perjes}. 
In matter, several authors have dealt with these solutions
\cite{bonnor60}$-$\cite{horvat}, showing new analysis 
and results. An interesting result that concerns us here was 
given by Das \cite{das62}, where it was shown 
that in the case of static charged
incoherent matter distributions the condition balance (\ref{majumdar2}), 
$\rho_{\rm e}=\rho$, implies the Majumdar-Papapetrou condition 
(\ref{majumdar1}), $V=\left(\sqrt{A}\pm\phi\right)^2$. This 
analysis was further developed in \cite{guilfoyle}. 
The other works dealt with a large
number of models of different charged bodies with several types of
shapes and mass density profiles which obey the Majumdar-Papapetrou
relation displaying new analysis and results.

Now, the multi-black hole Reissner-Nordstr\"om solutions have the
interesting property of being supersymmetric \cite{gibbonshull}, a
result that can be extended to all solutions belonging to the
Majumdar-Papapetrou system \cite{tod}. These solutions saturate
supersymmetric Bogomol'nyi bounds, are stable, and  may be
considered as ground states of the theory \cite{kallosh}.  These
Majumdar-Papapetrou solutions can also be embedded in supergravities
and superstring theories with dimensions higher than four. This has
been done for vacuum solutions where generalized extreme
Reissner-Nordstr\"om single \cite{tangherlini} and extreme multi-black
holes \cite{myers} have been studied, as well as brane extensions to
higher dimensions \cite{ivashchuck,peet}. The study of d-dimensional
solutions, with ${\rm d}\geq4$, 
in several theories is a hot topic. 
The higher dimensional
Kerr solutions were discussed in \cite{myersperry}. Higher dimensional
Weyl metrics were analyzed in \cite{bronnikov}.  There is now a
5-dimensional rotating black ring solution \cite{emparanreall} and its
charged generalization \cite{elvang} which extreme case belongs to the
Majumdar-Papapetrou class in 5-dimensions (see also
\cite{elvangemparan,harmark}). 
All the d-dimensional solutions, with ${\rm d}\geq4$, 
mentioned above are vacuum solutions.  It is then,
of course, important to analyze d-dimensional, ${\rm d}\geq4$, 
Majumdar-Papapetrou type
solutions in matter, a subject which we pursue here. 
One can ask whether lower dimensional theories, ${\rm d}<4$, 
admit Majumdar-Papapetrou solutions or not. As will be seen,  
${\rm d}=2$ and ${\rm d}=3$ yield singular expressions 
(either give zero or infinity) in our formulas for generic d 
Majumdar-Papapetrou systems. This is no surprise.
In ${\rm d}=3$ general relativity 
the analogues of Majumdar-Papapetrou solutions are pure point 
sources, since gravity does not propagate \cite{deser}. General relativity 
in ${\rm d}=2$ does not exist and so there is no analogue. On the 
other hand, for lower dimensional theories of gravity, other 
than general relativity, with extra fields such as a dilaton field, 
analogues of the Majumdar-Papapetrou solutions might be found, 
see \cite{mannohta} for a ${\rm d}=2$ system.  
Due to their singular behavior we avoid treating the lower-dimensional 
Majumdar-Papapetrou systems, and work for the ${\rm d}\geq4$ systems only.

In line with the beautiful paper of Majumdar
\cite{maj47} we will follow closely his analysis and render his
results in four dimensions into higher d-dimensions.  We will also
incorporate the interesting developments of Guilfoyle
\cite{guilfoyle}.  The plan of the present work is as follows:
Starting with a charged perfect fluid in d-dimensions, we impose a
Weyl type relation among the gravitational potential $V$ and the
electrostatic potential $\phi$, $V=V(\phi)$, obtain a relation among
the pressure and the charge and mass densities of the fluid, and show
that, for matter distributions that have a boundary where the
pressure is zero, only the case of charged dust
matter (zero pressure everywhere) has non-singular solutions.
This is done in Sects. \ref{model} to \ref{Vanishing}.  
In Sect. \ref{sectfunctional} solutions involving a relation between 
$g_{tt}$ and $\phi$ are analyzed. 
In Sect. \ref{sectexactsolutions} we show that 
the (d$-1$)-dimensional spacelike sub-manifold is conformal to a
Ricci-flat space only if the pressure vanishes.
In Sect. \ref{nature} the nature of the solutions are analyzed. 
In Sect.  \ref{Vanishing}  the vanishing of the material stresses 
is analyzed, showing also that if it is assumed that the
(d$-1$)-dimensional sub-manifold is conformal to a Ricci-flat space
then the pressure vanishes.
In Sect. \ref{sectweyl} an examination of generalized Weyl's 
axially symmetric solutions in d-dimensional spacetimes is done
for electrovacuum and in presence of charged matter. 
In Sect. \ref{sectboundary} a brief analysis of boundary value problems 
is made. 
Finally, in Sect. \ref{conclusions} we present final
comments and conclusions.

\section{The fundamental equations}
\label{model}

We write  the Einstein-Maxwell equations as ($c=1$)
\begin{eqnarray}
& &G_\munu= 8\pi 
\left( T_\munu+E_\munu\right)\, ,
\label{einst}\\
& & \nabla_\nu F^\munu = 4\pi J^\mu\,, \label{maxeqs}
\end{eqnarray}
where Greek indices $\mu, \nu$, etc., run from $0$ to
${\rm d}-1$. We have put $G_{\rm d}$, the ${\rm d}$-dimensional 
gravitational constant equal to one, $G_{\rm d}=1$, as well as $c=1$ 
throughout. Also, 
$g_\munu$ is the metric, $G_\munu=R_\munu-\frac{1}{2}g_\munu R$ is the
Einstein tensor, with $R_\munu$ being the Ricci tensor, and $R$ the Ricci
scalar. 
$E_\munu$ is the electromagnetic energy-momentum tensor, given by
\begin{equation}
4\pi E_\munu= {F_\mu}^\rho F_\nu{_\rho} -\frac{1}{4}g_\munu F_{\rho\sigma}
F^{\rho\sigma}\, ,\label{maxemt}
\end{equation}
where the Maxwell tensor is 
\begin{equation}
F_\munu = \nabla_\mu A_\nu -\nabla_\nu A_\mu\, , \label{ddemfield}
\end{equation} 
$\nabla_\mu$ being the
covariant derivative, and
$A_\mu$ the electromagnetic gauge
field. In addition, 
\begin{equation}
J_\mu = \rho_{\rm e}\, U_\mu\, ,\label{current}
\end{equation}
is the current density, $\rho_{\rm e}$ is the electric charge
density in the d-dimensional spacetime, 
and $U_\mu$ is the fluid four-velocity.
$T_\munu$ is the material energy-momentum tensor given by
\begin{equation}
T_\munu = \rho_{\rm m}\, U_\mu U_\nu + M_\munu\,,\label{fluidemt}
\end{equation}
where $\rho_{\rm m}$ is the fluid matter energy density in the
d-dimensional spacetime, and
$M_\munu$ is the stress tensor. Following Guilfoyle \cite{guilfoyle},
in our analysis we will use mainly a perfect fluid, in which case
$M_\munu$ is given by
\begin{equation}
M_\munu^{\rm perfect\,fluid}
= p\left( U_\mu U_\nu +  g_\munu\right)\,.\label{perfectf}
\end{equation}
Note that Majumdar \cite{maj47} uses mainly electrovacuum. 
Since the gravitational constant $G_{\rm d}$ has dimensions of
$\left({\rm length}\right)^{({\rm d}-3)}\times({\rm mass})^{-1}$, 
and we have set it to one,  $G_{\rm d}=1$, it implies 
that mass has dimensions of $({\rm length})^{({\rm d}-3)}$, while 
$\rho_{\rm m}$ has dimensions of $({\rm length})^{-2}$.

We assume the spacetime is static and that the metric can be written
in form
\begin{equation}
ds^2 = - V dt^2 + h_{ij}\,dx^idx^j\, ,\quad\quad i,j=1,....,{\rm d}-1\,, 
\label{metricdd}
\end{equation}
a direct extension of Eq. (\ref{metric4d}) to extra dimensions. 
The  gauge field and four-velocity are then given by
\beqa
& &A_\mu = \phi\,\delta_\mu^0\, ,\label{gauge1}\\
& &U_\mu =  -\sqrt{V\,}\, \delta_\mu ^0\, .\label{veloc1}
\eeqa
The metric spatial tensor $ h_{ij}$, the metric 
potential $V$ and the electrostatic potential $\phi$ are
functions of the spatial coordinates $x^i$ alone.

Initially, we are interested in the equations determing the metric potential
$V$ and the electric potential $\phi$. These are obtained respectively from the
$tt$ component of Einstein equations
(\ref{einst}) and from the $t$ component of Maxwell equations (\ref{maxeqs}). 
These equations give
\beqa
&& \partial_i\left(\sqrt{h\,}h^{ij}\partial_j V\right) = 
\frac{\sqrt{h\,}}{ 2V} h^{ij} \partial_i V\, \partial_j V 
+4\, \frac{{\rm d}-3} {{\rm d}-2}\sqrt{h\,}
\left[h^{ij}\partial_i\phi\partial_j\phi +4\pi V\left(\rho_{\rm m}+ 
\frac{{\rm d}-1}{{\rm d}-3}\,p\right) \, \right]\! ,\label{ttein}\\
&& \partial_i\left(\sqrt{h\,}h^{ij}\partial_j \phi\right)
=  \frac{1}{2V}\sqrt{h\,}h^{ij}\partial_i V\, \partial_j\phi+
4\pi\sqrt{h\,V\,} \rho_{\rm e}  \, , \label{max2}
\eeqa
where $h$ stands for the determinant of the metric $h_{ij}$, and
$\partial_i$ denotes the partial derivative with respect to the
coordinate $x^i$.
Notice that Maxwell equations (\ref{maxeqs}) imply just one equation,
the $t$ component, $\nabla_i F^{ti}=4\pi J^t$, showed in (\ref{max2}). 

Eqs. (\ref{ttein}) and (\ref{max2}) determine the
potentials $V$ and $\phi$ in terms of a set of unknown quantities.
Namely, the $({\rm d}-1)({\rm d}-2)/2$ spatial metric coefficients $h_{ij}$,
the fluid variables, energy density $\rho_{\rm m}$ and pressure $p$,
and the electric charge density $\rho_{\rm e}$.  There are exactly
$({\rm d}-1)({\rm d}-2)/2$ additional equations that come from the
Einstein equations, which in principle determine the $h_{ij}$ metric
components in terms of $\rho_{\rm m}$, $p$ and $\rho_{\rm e}$.  Hence,
to complete the system of equations it is necessary to provide the
energy and charge density functions, $\rho_{\rm m}$ and $\rho_{\rm
e}$, and also to specify the pressure $p$ or an equation of state for
the perfect fluid.
In the present analysis, we will not need the explicit form of
the space metric $h_{ij}$ and so the corresponding Einstein
equations will not be written here. Additional equations that 
can be used are the conservation
equations, $\nabla_\nu T^{\munu}=0$, which are sometimes useful 
in replacing a subset of Einstein's equations.
In the present case the conservation
equations yield
\beq
\partial_i\,p +\frac{1}{2V}\left(\rho_{\rm m}+p\right)\partial_i\, V
-\frac{1}{\sqrt{V\,}}\rho_{\rm e}\partial_i\,\phi =0\label{conserveq}.
\eeq
This is the relativistic analogous to the Euler equation, and
carries the information of how the pressure gradients balance the
equilibrium of the system.  In what follows we investigate some
particular cases of the above set of equations including
electro-vacuum, dust fluid, and a perfect fluid.

\section{Solutions involving a functional relation between 
$\bm g_{tt}\equiv V$ and $\bm\phi$}
\label{sectfunctional}

\subsection{The Equations}

We now assume the solutions of the d-dimensional spacetime 
to be of Weyl type where the metric
potential $g_{tt}\equiv V$ is a functional of the gauge potential $\phi$,
$V=V(\phi)$. Hence, Eqs. (\ref{ttein}) and (\ref{max2}) read,
respectively,
\beqa
&&\hskip -1.0cm\partial_i\!\left(\!\sqrt{h}h^{ij}\partial_j
\phi\right)\!\! =\!  \sqrt{h}\! \left(\!\frac{ V'}{2V}+ 
\frac{4}{V'}\frac{{\rm d}-3}{{\rm d}-2}- \frac{ V''}{V'}\right)h^{ij} 
\partial_i \phi \partial_j \phi +16\pi\,\sqrt{h}{V\over
V'}\left({{\rm d}-3\over {\rm d}-2}\,\rho_{\rm m}+ 
\frac{{\rm d}-1}{{\rm d}-2}\,p\right)\!\! , 
\label{ttein2} \\
 &&\hskip -1.cm \partial_i\!\left(\sqrt{h\,}h^{ij}\partial_j
\phi\right)\!\! = \frac{1}{2}\frac{V'}{V}\sqrt{h\,} h^{ij}\partial_i \phi\,
\partial_j\phi+ 4\pi\sqrt{h\,V\,} \rho_{\rm e} \, ,\hfill\label{max3}
\eeqa
where we have defined $V' = \displaystyle\frac{dV}{d\phi}$ and $V'' =
\displaystyle\frac{d^2 V}{d\phi^2}$.
Using the last two equations we get 
\beq 
\left({4}\,{{\rm d}-3\over {\rm d}-2}- {V''}\right)h^{ij}\partial_i\phi\,
\partial_j\phi -4\pi V'\sqrt{V\,}\rho_{\rm e} + {16\pi}\,V\left(
{{\rm d}-3\over{\rm d}-2}\,\rho_{\rm m}+ {{\rm d}-1\over {\rm d}-2}\,p\right)
=0\, . \label{cc1}
\eeq
Note the singular behavior of the 
lower dimensional cases, ${\rm d}=2$ and  ${\rm d}=3$, 
which will not be treated 
here.
Substituting the functional relation $V=V(\phi)$ into the
conservation equations (\ref{conserveq}) it follows
\beq
\partial_i p +\left[(\rho_{\rm m}+p){ V'\over 2{V\,}} -
 {\rho_{\rm e}\over\sqrt{V}} \right] \partial_i\phi=0 \, .\label{conserveq2}
\eeq
From this equation it is possible to show that the pressure $p$ turns out
to be a function of $\phi$ alone, and depends
only indirectly on the matter and charge densities, see Appendix. 
Hence, there is a relation of the form 
\beq
p = p(\phi),\label{relationpphi}
\eeq
among the pressure $p$ and the electric potential and $\phi$, or 
equivalently among $p$
and the metric potential $V$, i.e., $p = p(V)$.
A particular case of this relation appears in Sec. \ref{sectmpsol}. 
The basic system of equations to be solved is composed by Eqs. (\ref{max3}), 
(\ref{cc1}) and (\ref{conserveq2}).

\subsection{Electrovacuum solutions}
Here we generalize to d-dimensions, first the Weyl form and then the
Majumdar-Papapetrou form of electrovacuum solutions to d-dimensional
spacetimes. 

\subsubsection{Weyl form}
In vacuum one has $\rho_{\rm m}=0$, $\rho_{\rm e}=0$, and
$M_\munu=0$.  Using these in Eq. (\ref{cc1}) and assuming
$h^{ij}\partial_i\phi\, \partial_j\phi \neq 0$, it follows that
$\dspst{{4}\,{{\rm d}-3\over {\rm d}-2}-{ V''}=0}$, whose solution is
\beq
V(\phi)= A + B\,\phi + 2 \,{{\rm d}-3\over {\rm d}-2}\,\phi^2\,  
\, , \label{weylpotential}
\eeq
where $A$ and $B$ are arbitrary constants. These are Weyl type
solutions generalized from four to higher dimensions.  This result
holds for any spatial symmetry.  Thus, it generalizes Majumdar's
result, where he noticed that the function form (\ref{weylpotential})
with ${\rm d}=4$ would hold not only for the axial symmetry imposed by Weyl
\cite{weyl}, but also for any spatial symmetry.

\subsubsection{Majumdar-Papapetrou form}
The problem is further simplified by choosing $V(\phi)$ in the form of
a perfect square, as done by Majumdar \cite{maj47}. In the higher
dimensional case, this is accomplished by choosing $B=\pm
2{\sqrt{ 2\,A\,\frac{{\rm d}-3}{{\rm d}-2}}}$.  Therefore, the metric potential
$V$ reads
\beq 
V(\phi) = \left(\sqrt{A\,\,}\pm \sqrt{ 2 {{\rm d}-3\over 
{\rm d}-2}\,}\,\phi\right)^2
\, \hskip .2cm \, . 
\label{mppotential}
\eeq
Eq. (\ref{mppotential}) is the Majumdar-Papapetrou condition in 
d-dimensions for ${\rm d}\geq 4$. 

A word on the choice of the relation between the
potentials $V$ and $\phi$ is in order.
The constant $A$ can be normalized by some particular asymptotic
condition on the metric, or can be set to zero by a redefinition
of the electric potential $\phi$.
This can be done since the Einstein-Maxwell equations,
Eqs. (\ref{einst}) and (\ref{maxeqs}), depend upon $\phi$ only through
its derivatives, so we may remove the additive constant $A$
in Eq. (\ref{mppotential}) by performing the
transformation $\phi \longrightarrow \phi \mp \,\frac12
\frac{{\rm d}-2}{{\rm d}-3}\,A$,
and writing $V = 2\frac{{\rm d}-3}{{\rm d}-2} \,\phi^2$.

\subsection{Charged matter solutions}
\label{sectmpsol}

\subsubsection{Weyl form}
As mentioned above, the condition imposed on $V(\phi)$ to produce
Weyl type solutions in vacuum is $V''=4\,({\rm d}-3)/({\rm d}-2)$, see
Eq. (\ref{weylpotential}).  If the system has matter, we see
that, when the condition $V''=4\,({\rm d}-3)/({\rm d}-2)$ is
satisfied, Eq. (\ref{cc1}) holds only if
\beq 
\rho_{\rm e} = {4}\,{\sqrt{V}\over  V'}\left({{\rm d}-3\over {\rm d}-2}\,
\rho_{\rm m}+ {{\rm d}-1\over {\rm d}-2}\,p\right)\, .\label{weylfluid}
\eeq
This relation holds whether or not the potential $V$ is a perfect
square in $\phi$. This together with Eq. (\ref{weylpotential}) are the
Weyl type conditions valid for matter systems. Usually this condition
is not discussed in the literature, see however \cite{guilfoyle} for
the four-dimensional analysis. 

As in the electrovacuum case, lower dimensional spacetimes deserves special
attention. The Weyl condition in three dimensions follows by substituting
${\rm d}=3$ into Eqs. (\ref{weylpotential}) and (\ref{weylfluid}), what gives 
$$
\rho_{\rm e} = {4}\,{{\rm d}-1\over d-2}{\sqrt{A + B\phi}\over  B}\,p\, .
\label{weylfluid3d}
$$
This equation, together with Eqs. (\ref{conserveq}) and (\ref{max3}) indicates
that there can be found interesting Weyl type charged fluid solutions in 
three-dimensional spacetimes.

\subsubsection{Majumdar-Papapetrou form}

Now, we want to specialize from Weyl type solutions 
in matter to Majumdar-Papapetrou solutions in matter. 
We then look for particular solutions of Eq. (\ref{cc1}) by choosing
the metric potential $V$ in the Majumdar-Papapetrou form, 
i.e., in the form  of a perfect square as in Eq. 
(\ref {mppotential}). Substituting such a potential $V(\phi)$ into
(\ref{weylfluid}) it follows
\beq 
\rho_{\rm e} = \pm \sqrt{{2}\,{{\rm d}-3\over {\rm d}-2}}
\left(\rho_{\rm m}+ {{\rm d}-1\over {\rm d}-3}\,p\right)\, , \label{mpfluid}
\eeq
which is the generalized Majumdar-Papapetrou condition for matter in spacetimes
whose number of dimensions is ${\rm d}\geq 4$.
The plus sign goes with positive electric charge and the minus sign 
is chosen for negative electric charge.
This analysis is analogous to the four-dimensional case, see 
\cite{das62,gautreau,guilfoyle,bonnor80}.
The Majumdar-Papapetrou solution for $V$ in terms of  $\phi$ is then
complete, while the solution for $\rho_{\rm e}$ is given, as a function
of $\rho_{\rm m}$ and $p$, by Eq. (\ref{mpfluid}).

We now, following the interesting result of \cite{guilfoyle} in four
dimensions, show that the Majumdar-Papapetrou conditions rule out 
d-dimensional perfect fluid solutions with boundary surfaces on which
$p=0$, but $p\neq 0$ in the bulk. 
In order to see that, we substitute Eqs. (\ref{mppotential})
and (\ref{mpfluid}) and into  (\ref{conserveq2}) and obtain  
$({\rm d}-3)\,V\,\partial_i p = p\,\partial_i V$. This equation can easily be
integrated giving
\beq
p^{{\rm d}-3}= kV\,, \label{relationVp}
\eeq
where $k$ is an integration constant.
It follows from the last equation that the surface of zero pressure is also a
surface where the $g_{tt}$ coefficient of the metric vanishes, implying a
metric singularity. In the static spacetimes we are considering here,
the vanishing of $g_{tt}$ means infinite redshift, and such a kind
of surface is not allowed in the solution for a self-gravitating perfect 
fluid. Suppose a localized object such as a star, for instance. The surface
of the star is usually defined by imposing $p=0$ as a boundary condition. Eq. 
(\ref{relationVp}) implies that the surface of the star would be singular
(an infinite redshift surface), and the solution cannot represent a 
star. In cases like this, it is then necessary to take $k=0$, implying that
pure Majumdar-Papapetrou type stars do exist only for charged dust, 
i.e., for solutions in which $p=0$ everyhere, which is the case 
most treated in the literature \cite{bonnor60}-\cite{horvat}.
Of course, Majumdar-Papapetrou solutions without boundaries can have 
$p\neq0$ throughout matter. Another possibility 
is to consider Majumdar-Papapetrou solutions 
with some thin shell at the surface.

\section{A class of exact solutions}
\label{sectexactsolutions}

Here we show that the $({\rm d}-1)$-dimensional spatial section of the
spacetime satisfying the Majumdar-Papapetrou condition has Ricci
tensor proportional to the pressure of the fluid. If the pressure
vanishes the sub-space is conformal to a Ricci-flat Riemannian space. 

Let us then factor out a conformal factor in the 
$({\rm d}-1)$-space metric as follows,
\begin{equation}
h_{ij}\,dx^idx^j = {1\over W}\, \hat h_{ij}\,dx^i dx^j\, ,\label{spacemetric}
\end{equation}
where $W$ and $\hat h_{ij}$  are functions of the space coordinates 
$x^i$, $i=1,\,2,\, ...,\,{\rm d}-1$. 
There is no loss of generality in the above choice, since $W$  and
$\hat h_{ij}$ are arbitrary functions.

The next step is writing Einstein-Maxwell equations in 
terms of the potentials 
$V$ and $W$, and in terms of the tensors in the conformal space. 
The convention adopted in the following is that all quantities wearing hats
belong to the conformal
space and are associated to the conformal metric $\hat h_{ij}$.
For instance, the connection coefficients 
may be written as
$ \Gamma^i_{jk}= \hat\Gamma^i_{jk} -\displaystyle{{1\over 2W}
\left[\delta_k^i\hnabla _j W
+\delta_j^i\hnabla_k W - \hat h_{jk}\hnabla^i W\right], }$
where the connection coefficients 
$\hat\Gamma^i_{jk}$, and the covariant derivative $\hat \nabla_i$,
are built from the metric $\hat h_{ij}$. 

Now, the explicit form of the Ricci tensor in terms of $V$, $W$ 
and $\hat\Gamma^i_{jk}$ is needed.  After some algebra we obtain
\beqa
R_{tt} &=& -{W\over 2}\hnabla^2 V +{W\over 4V}\hnabla V\cdot\hnabla V
+{{\rm d}-3\over 4} \hnabla V\cdot\hnabla W\, ,\label{ricci_tt} \\
R_{ij}&=& \hat R_{ij} -{1\over 4V^2}\left(2V\hnabla_i\hnabla_j V -
\hnabla_i V\hnabla_j V\right) -{1\over 4VW}\left(\hnabla_i V \hnabla_j W 
+\hnabla_j V\hnabla_i W-{\hat h_{ij}}\hnabla V\cdot\hnabla W\, \right) 
       \nonumber \\
& & +{{\rm d}-3\over 4W^2}\left({2W}\hnabla_i\hnabla_j W -\hnabla_i
W\hnabla_j W  -{{\rm d}-1\over {\rm d}-3}\,\hat h_{ij}\hnabla 
W\cdot\hnabla W
+{2W\over {\rm d}-3}\,\hat h_{ij}\hnabla^2 W\right)\,  , \label{ricci_ij}
\eeqa
where the dot stands for the scalar product with respect
to the metric $\hat h_{ij}$.
It is then seen that the expressions for the components of the Ricci tensor 
are greatly simplified by choosing $V({\rm d}-3)\hnabla 
W =W\hnabla V $, which means
\beq
W = V^{1\over {\rm d}-3}\, . \label{conformalfactor}
\eeq
As argued previously, this choice can be done without loss
of generality since the metric $\hat h_{ij}$ is arbitrary.
Substituting (\ref{conformalfactor}) into (\ref{ricci_tt}) and
(\ref{ricci_ij}) it follows
\beqa
R_{tt} &=&   -{1\over 2\,}{V^{1\over {\rm d}-3}}\left(\hnabla^2 V - {1\over V}
\hnabla V\cdot\hnabla V\right)\, ,\label{ricci_tt_1} \\
R_{ij}&=& \hat R_{ij} -\frac14\,{{\rm d}-2\over {\rm d}-3,}{\hnabla_i V\over V}
{\hnabla_j V\over V}
+{1\over {\rm d}-3\, }{\hat h_{ij}\over 2V}\left(\hnabla^2 V- {1\over V}
\hnabla V\cdot\hnabla V\right) \,  . \label{ricci_ij_1}
\eeqa
Calculating the energy momentum tensor it is found
\beqa
& &G_{tt} -{g_{tt}\over {\rm d}-2}G= -2{{\rm d}-3\over {\rm d}-2\,}
V^{1\over {\rm d}-3\,}\,\hnabla \phi\cdot \hnabla \phi 
+ 8\pi V{{\rm d}-3\over {\rm d}-2}\,
\left(\rho_{\rm m} +{{\rm d}-1\over {\rm d}-3}\, p\right)\, , 
\label{semt_tt}\\
& &G_{ij}  - {g_{ij}\over d-2}G
= -2  \hnabla_i \phi \hnabla_j \phi +{2\over {\rm d}-2} \hat h_{ij}
\nabla \phi \cdot\nabla \phi  +{8\pi\over {\rm d}-2}{\hat h_{ij}
\over V^{1\over {\rm d}-3}}\left(\rho_{\rm m}-p\right) \, . \label{semt_ij}
\eeqa
{From} the $tt$ component of Einstein equations,
and Eqs. (\ref{ricci_tt_1}) and (\ref{semt_tt}), 
we obtain the following equation for $V$  
\beq
\hnabla^2 V-{1\over V}\hnabla V\cdot\hnabla V
= 4{{\rm d}-3\over {\rm d}-2\,}{1\over V}\,
\hnabla\phi\cdot \hnabla \phi +16\pi \,{{\rm d}-3\over {\rm d}-2}
\,{1\over V^{1\over {\rm d}-3}}
\left(\rho_{\rm m}+\frac{{\rm d}-1}{{\rm d}-3}\,p\right) \, .\label{poissoneq1}
\eeq
Additionally, using Eqs. (\ref{ricci_ij_1}) and (\ref{semt_ij}), and
the Einstein equations, it follows
\beqa
\hat R _{ij} &=&\frac14\,{{\rm d}-2\over {\rm d}-3}{\hnabla_i V\over V}{\hnabla_j 
V\over V}
-\frac12\,{1\over {\rm d}-3\, }{\hat h_{ij}\over V}\left(\hnabla^2 V- {1\over V}
\hnabla V\cdot\hnabla V\right) \, \nonumber \\ 
& &-2\hnabla_i\phi\,\hnabla_j\phi
+{h_{ij}\over {\rm d}-2}\left[2\hnabla\phi\cdot\hnabla\phi+{8\pi
\over V^{1\over {\rm d}-3}}\left(\rho_{\rm m} -p\right)\right] \, .
\eeqa
Comparing the last two equations we get
\beq 
\hat R_{ij} =\frac14{{\rm d}-2\over 
{\rm d}-3}{\hnabla_i V\over V} {\hnabla_j V \over V}
-2\hnabla_i\phi\,\hnabla_j\phi 
-{16\pi\over {\rm d}-3}\, {p\over V^{1\over {\rm d}-3}}
 \,\hat h_{ij}\, \label{spacericci} .
\eeq

We also need to rewrite the Maxwell equation (\ref{max2}) under the
choices given by Eqs. (\ref{spacemetric}) and (\ref{conformalfactor}).
The resulting equation for the electric potential $\phi$ is then
\beq
\hnabla^2\phi = {\hat \nabla V\over 2V}\cdot \hat \nabla\phi+
4\pi{1\over V^{1\over {\rm d}-2}} \rho_{\rm e}  \, , \label{maxeqs3}
\eeq

Once the energy density, pressure and charge density are specified,
the system of equations formed by (\ref{poissoneq1}),
(\ref{spacericci}) and (\ref{maxeqs3}) supply all the equations needed
to determine the variables $V$, $\hat h_{ij}$ and $\phi$.
Eq. (\ref{poissoneq1}) can be thought as the equation that furnishes
$V$, and Eqs. (\ref{spacericci}) determine the inner metric $\hat
h_{ij}$, while the Maxwell equation (\ref{maxeqs3}) determines the
electric potential $\phi$.

If we  assume the functional relation among $V$ and $\phi$ is
given by the Weyl type potential (\ref{weylpotential}), then the
number of unknown variables is reduced by one. The same happens to the
number of equations because, in such a case, Eqs. (\ref{poissoneq1})
and (\ref{maxeqs3}) become identical, and the fluid variables
$\rho_{\rm m}$, $p$ and $\rho_{\rm e}$ are connected by relation
(\ref{weylfluid}).
If we  further 
assume that the relation between $V$ and $\phi$ is the 
Majumdar-Papapetrou condition (\ref{mppotential}), Eqs. 
(\ref{spacericci}) read
\beq
\hat R _{ij} = -16\pi{p \over V^{1\over {\rm d}-3}} \hat h_{ij} \, .
\label{spacericci2}
\eeq
A further conformal transformation such that $\hat h_{ij} = 
V^{1\over {\rm d}-3}\bar h_{ij}/p$
leads to a new Ricci tensor $\bar R_{ij}=-16\pi\bar h_{ij}$,
which is characteristic of a $({\rm d}-1)$-space  of constant  
scalar curvature, $\bar R = {\rm constant}$.
 The static spacetime fulfilled by
a charged fluid whose fields satisfy the Majumdar-Papapetrou
condition has a $({\rm d}-1)$-dimensional spacelike sub-manifold
which is conformal to a space of constant curvature. When the 
pressure vanishes, the curvature is zero and the sub-manifold
is conformal to a Ricci-flat space. 
These results are independent of the spatial symmetry of the
matter distribution. 

Defining now 
\beq
U = \frac{1}{\sqrt{V}}\,, \label{redefiningV}
\eeq
the metric of 
the spacetime in the presence of a generic 
charged fluid can be put into the form
\beq
ds^2 = -U^{-2}dt^ 2 + U^{2\over {\rm d}-3}\hat h_{ij}dx^i dx^j \, .
\label{mpmetric}
\eeq
Using this definition of $U$ and Eq. (\ref{poissoneq1}),  one obtains a
Poisson-like equation for $U$,
\beq
\hnabla^2U = -\,8\pi\, {{\rm d}-3
\over {\rm d}-2}\,U^{{\rm d}-1\over {\rm d}-3}
\left(\rho_{\rm m}+\frac{{\rm d}-1}{{\rm d}-3}\,p\right) \, .
\label{poissoneq}
\eeq
In the particular case of 
vacuum, the last equation  reduces to the Laplace equation, just as in
four dimensions \cite{maj47}. Moreover, using Eqs. (\ref{mppotential}),
the relation between $U$ and $\phi$ results
\beq
U=\frac{1}{\sqrt{A\,} \pm \
\sqrt{2\frac{{\rm d}-3}{{\rm d}-2}\,}\,\,\phi} \, .
\label{relationUphi}
\eeq
A spacetime metric in the form (\ref{mpmetric})
was used in Ref. \cite{myers}, as an ansatz, to study 
d-dimensional Majumdar-Papapetrou vacuum solutions.

\section{Nature of the solutions}
\label{nature}

The class of exact solutions discussed in the previous section
correspond to static spacetimes whose metric coefficient $g_{tt}$ is a
special function of the electric potential $\phi$, $g_{tt}=V=\frac{1}{U^{2}}$
with $U$ and $\phi$ related by (\ref{relationUphi}).  Note, however,
that there is some arbitrariness on the choice of the relation between
the potentials $U$ and $\phi$. Two interesting possibilities worth to
be mentioned here.
First,  the additive constant $A$ can be made equal to zero 
by performing the
transformation $\phi \longrightarrow \phi \mp 
A\sqrt{\frac12\,\frac{{\rm d}-2}{{\rm d}-3}}$, and
writing ${\sqrt V} = \frac{1}{U}=2\,\frac{{\rm d}-3}{{\rm d}-2}\,\phi^2$ 
(see also the discussion
at the end of Sect. \ref{sectexactsolutions}).
Second, if $A\neq 0$, a re-parameterization of the time coordinate 
of the
form $t \longrightarrow t/\sqrt{A\,}$ transforms the $g_{tt}$ 
coefficient into the form 
$g_{tt} = \frac{1}{U^{2}}=\left(1 \pm 
\sqrt{2A\frac{{\rm d}-3}{{\rm d}-2}\,}\phi\right)^2\,.$
Furthermore, since  $A$ is an arbitrary integration
constant, it can then be chosen appropriately according to 
the choice of the
units of electric charge.  
That is to say, one may
choose $A$ such that $2A\,\frac{{\rm d}-3}{{\rm d}-2}=1 $, 
which implies 
$ \frac{1}{U} = 1\pm  \phi\,$
in appropriate units.

In the Newtonian limit one has $U \simeq 1 + \varphi$,
 and Eq. (\ref{poissoneq}) reduces to the Poisson equation for the
gravitational potential $\varphi$, $\hat\nabla^2\varphi=-8\pi\,
\frac{{\rm d}-3}{{\rm d}-2}\,\rho_{\rm eff}$,
 with the effective Newtonian matter density given by 
$\rho_{\rm eff}=\rho_{\rm m}+\frac{{\rm d}-1}{{\rm d}-3}\,p$.
It also follows that 
$\varphi=\mp \phi\,.$
As in four dimensions, the d-dimensional spacetime fulfilled with a 
charged fluid satisfying the Majumdar-Papapetrou condition is the
analogous to a d-dimensional Newtonian system of charged 
self-gravitating fluid in static equilibrium. In fact, the quantity
 $\rho_{\rm eff}= \rho_{\rm m}+\frac{{\rm d}-1}{{\rm d}-3}\,p$
is the effective energy density acting as source of the gravitational
field, and it can be thought as the Newtonian matter density. 
The effect of the dimensionality 
of the spacetime on the effective energy density is in the sense of
diminishing the weight of the pressure, as $\rho_{\rm eff}$ 
varies from $\rho_{\rm m}+3p$
for ${\rm d}=4$ to $\rho_{\rm m} +p$ in the limit 
${\rm d}\longrightarrow\infty$.
Additionally, the factor $\sqrt{2\,\frac{ {\rm d}-3}{{\rm d}-2}}$
also depends on the dimension of the spacetime.
However,  this factor  can be made equal to unity by 
choosing appropriate units. In fact, by putting back the 
gravitational constant of gravitation ($G_{\rm d}$) into Eq. (\ref{mpfluid}), 
the factor $ \sqrt{2\frac{{\rm d}-3}{{\rm d}-2}}$   is 
replaced by $\sqrt{2\frac{{\rm d}-3}{{\rm d}-2}\, 
G_{\rm d}\,}$. Then, we may choose
units such that 
$2\,\frac{{\rm d}-3}{{\rm d}-2}\,G_{\rm d}=1$, yielding 
$\frac{\rho_{\rm e}}{\rho_{\rm eff}} =\pm 1$ for all $d>3$. 
This is in accordance to the fact that what really matters for the
balancing of the gravitational and electromagnetic forces is the
relation between $g_{tt}$ and $\phi$ being the same at every point
of spacetime, as Eq. (\ref{mppotential}) shows.
In other words, if the charge and the effective energy densities bear
the same constant of proportionality, the system will be in static
equilibrium owing to the balancing of electric and gravitational
forces.  The value of such a proportionality parameter depends on the
system of units one chooses, and this choice is, of course, dependent
upon the dimensionality of the spacetime.

\section{Vanishing of the material stresses for a proper choice
of the internal field}
\label{Vanishing}

In three dimensions the Ricci tensor is proportional to the Riemann
tensor, so the three-space Ricci tensor $\hat R_{ij}$ is proportional to
the 3-space Riemann tensor $\hat R_{ijkl}$. Now, when $p=0$,
Eq. (\ref{spacericci2}) shows that the spatial Ricci tensor is zero,
$\hat R_{ij}=0$, so that the Riemann tensor is also zero, and the
inner spatial three-space is flat. The converse is also true, i.e.,
if the inner three-space is flat then $p=0$.  This was shown by Majumdar
\cite{maj47}.

In higher dimensions, however, the proportionality between Ricci and
Riemann tensors is not valid, so the vanishing of the pressure does
not imply a conformally flat (d$-1$)-space.  On the other hand, if we
assume that the metric is of the form (\ref{mpmetric}) and assume
further that the internal metric is Euclidean, $\hat h_{ij} =
\delta_{ij}$, then $\hat R _{ij}$ are identically zero and
Eqs. (\ref{spacericci2}) imply that the pressure vanishes.  This can
also be done for more general material stresses other than perfect
fluid pressures, partially generalizing thus Majumdar's result for four
dimensions.

\section{Examination of generalized Weyl's axially symmetric solutions
in higher dimensions}
\label{sectweyl}

A further study appearing in Majumdar's paper \cite{maj47} 
is a discussion and analysis of Weyl's results \cite{weyl} 
on axisymmetric solutions. 
Majumdar showed that when the relation between the metric potential 
and the electric potential is of the form of Eq. (\ref{weylcondition}), 
then there is no need of requiring axial symmetry. 
The original works by Weyl \cite{weyl} and Majumdar \cite{maj47} 
analyzed axisymmetric solutions mainly in electrovacuum 
case. The problem of pure gravitational field inside uncharged matter
(with no electromagnetic field) was  considered by Majumdar 
without assuming any particular kind of matter.
The case inside charged matter was only touched upon by Majumdar
in Sect. VII B of his paper \cite{maj47}. 
In the spirit of the present work, we extend the analysis
of axisymmetric Weyl solutions to the case of higher dimensions 
both in vacuum and 
with charged matter. As a particular case we discuss the 
four-dimensional case in charged matter, somehow completing 
Majumdar's discussion. 
The main aim of this section is then to find the explicit form of the
equations for the metric potentials and for the electrostatic potential
in the axisymmetric Weyl form.
The special cases where the metric potential $V$ assumes, as function of
the electric potential $\phi$, the Weyl and the Majumdar-Papapetrou
forms are depicted separately.
This is, in certain sense, an example of what was found in the previous
sections, particularized to the Weyl axisymmetric form of the metric 
in a d-dimensional spacetime. 

\subsection{The Equations}

In four dimensions for the Weyl axisymmetric metric one usually uses, instead
of $V$ in Eq. (\ref{metric4d}), the potential $\mu_0$ such that the potential 
$V$ appearing in Eq. (\ref{metric4d}) is $V={\rm e}^{2\mu_0}$, 
where $\mu_0$ is a function of $r$ and $z$ (see also
Sect. \ref{sectexactsolutions}). There are two other Weyl  potentials, 
one is $\mu_1$ 
which is usually related
to the Weyl radial coordinate $r$, as $r={\rm e}^{2\mu_1}$, the 
other is $\nu$ which is generally a function of $r$ and $z$. 
Then, the four-dimensional metric in Weyl axisymmetric form is written 
in terms of the functions $\mu_0$, $\mu_1$, and $\nu$, as
$ds^2={\rm e}^{2\mu_0}(- dt^2 +
{\rm e}^{2\mu_1}d\varphi^2)+ e^{2\nu}
\left(dr^2+dz^2\right)$, or putting ${\rm e}^{2\mu_1}=r$, as 
$ds^2= {\rm e}^{2\mu_0}\,(-dt^2 + r^2\,d\varphi^2)+
e^{2\nu} \left(dr^2 + dz^2\right)$,
where $(t,\varphi,r,z)$ are spacetime 
cylindrical type coordinates. The Einstein field equations 
can then be obtained and analyzed 
in terms of the Weyl coordinates. 
Interestingly enough this can also be done in higher dimensions. 
The axisymmetric metric for higher dimensions is given by Emparan 
and Reall \cite{emparanreall} (see also \cite{bronnikov} and references 
therein for different higher dimensional generalizations of Weyl 
axisymmetric solutions). 
They assumed that the spacetime has ${\rm d}-2$ non-null Killing vectors 
in which case the metric can be put into the form
\beq
ds^2= -{\rm e}^{2\mu_0} dt^2 + 
          {\rm e}^{-\frac{2\mu_0}{{\rm d}-3}}\,\,
           \sum_{i=1}^{{\rm d}-3}{\rm e}^{2\mu_i}(dx^i)^2 +
          {\rm e}^{2\nu} \left(dr^2+dz^2\right)  \, ,
\label{weylmetricdd} 
\eeq
where the functions $\mu_0$, $\mu_i$ and $\nu$ depend upon the
coordinates $r$ and $z$ only.
The function $\mu_0$ again plays the role of the gravitational potential.
The  Latin index $i$ runs from $1$ to ${\rm d}-3$ and the coordinates $x^i$
label ${\rm d}-3$ spatial dimensions of the spacetime,  while the two 
remaining 
spatial dimensions are labelled as $x^{{\rm d}-2}=r$ and $x^{{\rm d}-1}=z$. 
The metric (\ref{weylmetricdd}) satisfies the constraint 
(see e.g. Ref. \cite{harmark}),
\beq
\exp\left(\sum_{i=1}^{{\rm d}-3} \mu_i\right) = r\, . \label{constraintdd}
\eeq
Such a constraint implies that the function $\Phi$, defined by
$\Phi= \sum_{i}^{{\rm d}-3}\mu_i$, is harmonic 
\beq
\nabla^2 \sum_{i=1}^{{\rm d}-3}\mu_i = 0\, ,      \label{laplaceaxisdd}
\eeq 
i.e., it satisfies a Laplace equation. 
It was found that, in vacuum, Einstein equations imply that the
functions $\mu_0$, and $\mu_i$ satisfy a Laplace equation 
in a flat metric, $\nabla^2 \mu_\alpha =0\, ,$
where $\nabla^2$ is the Laplacian operator in d dimensions, 
and $\alpha$ runs from $0$ to ${\rm d}-3$, while $\nu$ is given 
as a function of $\mu_0$ and $\mu_i$.
Some particular solutions have been 
found in the vacuum case for ${\rm d}=5$, and also in higher dimensions
(see \cite{harmark} and references therein). Moreover, the black ring
solution of Emparan and Reall \cite{emparanreall} was generalized to include
electric charge \cite{elvang}. 
We then consider the generalized {\rm d}-dimensional static axisymmetric Weyl
spacetimes, whose metric is written as in Eq. (\ref{weylmetricdd}),
and investigate the general properties of the solutions in the presence of
charged matter and in electrovacuum.
We assume the  {\rm d}-dimensional spacetime inside matter satisfies the 
following
conditions: (i) there exists a line element of the form (\ref{weylmetricdd}),
where the functions $\mu_i$ satisfy the constraint (\ref{constraintdd}),
(ii) the matter content is given by an energy-momentum tensor of the form
(\ref{fluidemt}), with $M_\munu$ given by (\ref{perfectf}), (iii) the metric
potential $g_{tt}= {\rm e}^{2\mu_0}$ is connected to the electric potential
$\phi$ by (\ref{weylpotential}). Then,
using the metric in the form (\ref{weylmetricdd}) and the constraint
(\ref{constraintdd}), the Einstein equations yield
\beqa
&&\nabla^2 \mu_0 = 16\pi\,
\frac{{\rm d}-3}{{\rm d}-2}\, {\rm e}^{2\nu}
\left[\rho_{\rm m} + \left(\frac{{\rm d}-1}{{\rm d}-3}\right)p\right] +
4\,
\frac{{\rm d}-3}{{\rm d}-2}\,{\rm e}^{-2\mu_0}\,
\left[\left(\partial_r\phi\right)^2+
\left(\partial_z\phi\right)^2\right]\, , \label{weyleinst00dd}\\
&&\nabla^2 \mu_i = 32\pi \frac{{\rm e}^{2\nu}}{{\rm d}-3}\, p\,  ,
 \label{weyleinstiidd}\\
&&\partial_r\nu  = -\frac{1}{2r} - \frac{1}{{\rm d}-3}\partial_r \mu_0+
\frac{r}{2}
 \sum_{i=1}^{{\rm d}-3} \left[\left(\partial_r \mu_i\right)^2 
        -\left(\partial_z \mu_i\right)^2\right] \nonumber\\ 
&&  \quad\quad\quad +\frac{r}{2}\left[ \frac{{\rm d}-2}{{\rm d}-3} 
 \left[\left(\partial_r \mu_0\right)^2-\left(\partial_z \mu_0\right)^2\right] 
- 2{\rm e}^{-2\mu_0}\,\left[\left(\partial_r \phi\right)^2 
-\left(\partial_z \phi\right)^2\right]\right] \, , \label{weyleinstrrzzdd}\\
&&\partial_z \nu  = -\frac{1}{{\rm d}-3}\partial_z \mu_0 
+r\sum_{i=1}^{{\rm d}-3}
\partial_r \mu_i\,\partial_z \mu_i +
r\left[\frac{{\rm d}-2}{{\rm d}-3}\partial_r \mu_0\,
\partial_z \mu_0 - 2{\rm e}^{-2\mu_0}\,\partial_r\phi\, 
\partial_z\phi\right]\, , 
         \label{weyleinstrzdd}
\eeqa
where the Latin index $i$ runs form $1$ to ${\rm d}-3$. The above equations can
also be obtained from Eqs. (\ref{poissoneq}) and (\ref{spacericci})
with the obvious identifications $V= {\rm e}^{2\mu_0}$, $\hat h_{ij}=
{\rm e}^{2\mu_i}$ for $i=1$ to ${\rm d}-3$, and for the other two spacelike
coordinates, $x^{({\rm d}-2)}=r$ and $x^{({\rm d}-1)}$,
$\hat h_{rr}= \hat h_{zz} = {\rm e}^{2[\nu - \mu_0/({\rm d}-3)]}$.
The solely Maxwell equation is
\beq
\nabla^2\phi = 4\pi\, {\rm e}^{2\nu+\mu_0} \rho_{\rm e} +2\partial_r\phi\, 
\partial_r \mu_0 + 2\partial_z\phi\,\partial_z \mu_0\, , \label{weylmaxwelldd}
\eeq
which has exactly the same form as in four-dimensional spacetime.

Note that the special form of the metric (\ref{weylmetricdd}) implies
that the pressure vanishes. For adding Eq. (\ref{weyleinstiidd}) over
$i$, from $i=1$ to $i={\rm d}-3$, one finds $\nabla^2\sum_{i=1}^{{\rm
d}-3}\mu_i =32\pi\, {\rm e}^{2\nu}\, p$, and comparing to
(\ref{laplaceaxisdd}) gives $p=0$.  This result is consistent with
what was shown in Sect.  \ref{sectexactsolutions}.  The choice of the
metric of the {\rm d}-dimensional spacetime in the axisymmetric form
(\ref{weylmetricdd}) together with the Weyl condition
(\ref{weylpotential}) implies the vanishing of the pressure of the
perfect fluid.  It is straightforward to show that this is true for
any kind of matter, not only for a perfect fluid, i.e., in the present
conditions, all the material stresses $M_{ij}$ vanish.
In the following we investigate some general properties of the solutions
to the above system of equations.

\subsection{Electrovacuum solutions}
Consider first the electrovacuum case. The equations governing the 
metric and electric potentials are obtained from the system 
(\ref{weyleinst00dd})-(\ref{weylmaxwelldd}) with $\rho_{\rm m}=0$,
$p=0$, and $\rho_{\rm e}=0$. Let us consider the 
Weyl and Majumdar-Papapetrou cases separately.

\subsubsection{Weyl form}
In the d-dimensional case, the Weyl form of the relation between the metric
coefficient $g_{tt}$ and the electric potential $\phi$ is  
$ g_{tt}= {\rm  e}^{2\mu_0} = A + B\phi + 2{\rm d -3\over d-2}\phi^2$,
with $B\neq \pm 2\sqrt{2{\rm d -3 \over d-2}{A\,}}$ (see also Eq. 
(\ref{weylpotential})). Following Weyl \cite{weyl}, and defining an auxiliary
function $\chi$ by
\beq
\chi = \int{{\rm d}\phi\over A +B\phi + 2{\rm d -3\over d-2}\phi^2} =
{2\over \sqrt{B^2-8{\rm d -3\over d-2}\,A\,}}\tanh^{-1}
\left(4{\rm d -3\over d-2}\phi +B
\over\sqrt{B^2-8{\rm d -3\over d-2}\,A\,}\right)\, ,\label{chidefdd}
\eeq
it is possible to show that Eqs. (\ref{weyleinst00dd}) and 
(\ref{weylmaxwelldd}) become identical, and assume the Laplacian form,
i.e., 
\beq
\nabla^2 \chi =0\, . \label{chiequation0}
\eeq
For ${\rm d}=4$, Eq. (\ref{chidefdd}) reduces to the 
result presented by Majumdar \cite{weyl}, 
$\chi = -\int {{\rm d}\phi\over{A +B\phi+\phi^2}}= {2\over\sqrt{B^2-4A^2\,}}
\tanh^{-1}\left(2\phi+B\over{\sqrt{B^2-4A^2\,}}\right)$.
The properties of the other metric potential $\nu$ are analyzed in
detail by Majumdar \cite{maj47} and will not be considered here.

\subsubsection{Majumdar-Papapetrou form}

The particular  Majumdar-Papapetrou form happens when 
$B=\pm 2\sqrt{2{\rm d-3\over d-2}{A}}$, so that 
${\rm  e}^{2\mu_0}=\left(\sqrt{A\,} \pm\sqrt{2\frac{{\rm d}-3} 
{{\rm d}-2}\,}\phi\right)^2\, .$
With this, the relation among $\chi$ and the
electric potential $\phi$ results very simple,
\beq
\chi = \sqrt{\frac12\,{\rm d-2\over d-3}\,}
{1\over \sqrt{A\,}\pm\sqrt{2\frac{{\rm d}-3}{{\rm d}-2}\,}\phi} \, ,
\label{chidefddmp}
\eeq
with $\chi$ satisfying the Laplace equation (\ref{chiequation0}),
which is obtained form  (\ref{weyleinst00dd}), or from
(\ref{weylmaxwelldd}).
Notice that
$\chi$ is proportional to the function $U$ introduced in Sect.
\ref{sectexactsolutions} (see Eq. (\ref{relationUphi})).

It is interesting to study here the four-dimensional case. 
Substituting ${\rm d}=4$ into Eq. (\ref{chidefddmp}) it gives  
$\chi = {1\over \sqrt{A\,} \pm\phi}\,$.
Hence, in the four-dimensional Majumdar-Papapetrou case,
the potential $\chi$ is connected to the metric potential by  
the well known relation $\chi = {1\over \sqrt{V}}={\rm e}^{-\mu_0}$, as
found by Majumdar \cite{maj47} and Papapetrou \cite{papa47}.
Moreover, we find, for instance,  $\partial_r \,\mu_0\partial_z
\mu_o=e^{2\mu_0}(\sqrt{A}\pm\phi)^2\partial_r\phi\,\partial_z\phi$, etc.
Therefore, imposing the  Majumdar-Papapetrou condition we get
$\partial_r \nu = -\partial_r \mu_0$ and $\partial_z\nu = -\partial_z \mu_0$,
which yields
\beq
\nu = -\mu_0 + {\rm constant}\, .\label{mpaxisymmetric4dnew}
\eeq
The four-dimensional metric is then of the form
$ds^2= {\rm e}^{2\mu_0}\,dt^2 + {\rm e}^{-2\mu_0}\,
\left(dr^2 + r^2 d\theta^2
+ dz^2\right)$, where the spatial section of the spacetime is conformal
to a flat space. The inner (three-dimensional) metric is Euclidean,
in accordance to the results of Sect. (\ref{sectexactsolutions}) for
spacetimes with matter without stresses and satisfying the 
Majumdar-Papapetrou condition. The stresses are zero here due to the
a priori chosen axial symmetry of the metric.

On the other hand, differently from the four-dimensional case, 
in d dimensions there is no 
relation between $\nu$ and $\mu_0$ such as Eq. (\ref{mpaxisymmetric4dnew}).
This can be seen  by substituting the Majumdar-Papapetrou condition
into Eqs. (\ref{weyleinstrrzzdd}) and (\ref{weyleinstrzdd}), what gives
\beqa
&&\partial_r\nu  = -\frac{1}{2r} - \frac{1}{{\rm d}-3}\partial_r \mu_0
+\frac{r}{2}\sum_{i=1}^{{\rm d}-3} \left[\left(\partial_r \mu_i\right)^2 
-\left(\partial_z \mu_i\right)^2\right] \, , 
\label{weyleinstrrzzddmp}\\
&&\partial_z \nu  = -\frac{1}{{\rm d}-3}\partial_z \mu_0 
+r\sum_{i=1}^{{\rm d}-3} 
\partial_r \mu_i\,\partial_z \mu_i \, .
\label{weyleinstrzddmp}
\eeqa
Therefore, in the general {\rm d}-dimensional Weyl axisymmetric spacetime,
the Majumdar-Papapetrou condition (\ref{mppotential}) does not imply any
further special restriction to the space metric other than the Weyl
condition (\ref{weylpotential}) does. This result is related to the fact
that in {\rm d}-dimensional spacetimes the Majumdar-Papapetrou condition
does not imply that the metric of the inner space, $\hat h_{ij}$ (see Sect. 
\ref{sectexactsolutions}),  for vanishing matter stresses, is necessarily
flat, as it does in the case of four-dimensional spacetimes.

\subsection{Charged matter solutions}

The main properties of the solutions inside charged matter are analyzed 
in this section.

\subsubsection{Weyl form}

As shown in Sect. \ref{sectfunctional}, if we impose 
the Weyl relation $\displaystyle{{\rm e}^{2\mu_0}= A + B\phi + 
2\frac{({\rm d}-3)}{{\rm d}-2}\phi^2}$, then Eq. (\ref{weyleinst00dd}) 
become identical
to (\ref{weylmaxwelldd}) if the following equation holds
(see Eq. (\ref{weylfluid}))
\beq 
\rho_{\rm e} = {2}{{\rm d}-3\over {\rm d}-2}
{{\rm e}^{-\mu_0}\over{\mu_0}^\prime}\,\rho_{\rm m}=
{2}{{\rm d}-3\over {\rm d}-2} {\sqrt{A + B\phi + 
2\frac{({\rm d}-3)}{{\rm d}-2}\phi^2\;}\over B 
+4\frac{({\rm d}-3)}{{\rm d}-2}\phi}\,\rho_{\rm m}\, , 
\label{axisweylfluid}
\eeq
where the prime stands for the total derivative with respect to
$\phi$. In ${\rm d}=4$ dimensions one obtains 
$\rho_{\rm e}=\pm 2\frac{\sqrt{ A +B\phi+\phi^2}}{B+2\phi}\,
\rho_{\rm m}\,$. From Eq. (\ref{axisweylfluid}) it is
obtained the resulting equation for the potential $\chi$ defined
by (\ref{chidefdd}). It is
\beq
\nabla^2\chi = -8\pi\,{\rm d-3\over d-2}\,{\rm e}^{2\nu}\,
{\sqrt{{B^2\over 4} -2{\rm d-3\over d-2}\,A\,}\over\tanh\left(
\sqrt{{B^2\over 4}-2{\rm d-3\over d-2}\,A\,}\,\chi\right)}\, \rho{_m}\, .
\eeq
Thus, the general properties of the
axisymmetric solution satisfying the Weyl condition (\ref{weylpotential})
are verified as expected. The particular four-dimensional case considered
by Weyl \cite{weyl} and Majumdar \cite{maj47} follows from the above
analysis by choosing ${\rm d} =4$, and noting that the energy density 
function $\sigma$ used by Majumdar is related to the scalar invariant energy
density $\rho_{\rm m}$ by $\sigma={\sqrt{{B^2\over 4} -A\,}\,\,{\rm e}^{2\nu}
\over\tanh\left(\sqrt{{B^2\over 4}-A\,}\,\chi\right)}\, \rho_{\rm m}$.

\subsubsection{Majumdar-Papapetrou form}

The Majumdar-Papapetrou relation is now ${{\rm  e}^{\mu_0}=\left(\sqrt{A\,} 
\pm \sqrt{2\frac{{\rm d}-3}{{\rm d}-2}\,}\phi\, \right)}$, and
it is straightforward showing that Eq. (\ref{axisweylfluid}) reduces to
(\ref{mpfluid}) but here with $p=0$, i.e., $\rho_{\rm e} =
\sqrt{{2}{{\rm d}-3\over {\rm d}-2}\,}\rho_{\rm m}$.
As in the electrovacuum case,
the potential $\chi$ which satisfies the Majumdar-Papapetrou
condition given by
(\ref{chidefddmp}), is proportional to $U$,
where $U= {\rm e}^{-\mu_0} $ is the potential defined in
Eq. (\ref{relationUphi}).
In addition, the resulting equation for $ V={\rm e}^{2\mu_0}$ 
(or for $\phi$), obtained from (\ref{weyleinst00dd})
[or (\ref{weylmaxwelldd})], can be written in terms of the potential $\chi$ and
 put into the form (\ref{poissoneq}). Namely,
 \beq
\nabla^2\chi = -8\pi\,{\rm d-3\over d-2}\,{\rm e}^{2\nu}\,\chi\, \rho_{\rm m}\, .
\eeq
Hence, the spacetime metric can be written in Majumdar-Papapetrou form
(\ref{mpmetric}), in which the  metric of the inner space,   $\hat h_{ij}$,
is the flat metric. 
In four dimensions we have ${\rm e}^{2\nu}= {\rm e}^{-2\mu_0} = \chi^2$, and
 the potential $\chi$  satisfies the equation $
\nabla^2 \chi = -4\pi\, \chi^{3} \rho_{\rm m} \,$, as found in Refs.
\cite{maj47,papa47}. 
The same comments with respect to the function $\nu$
made at the end of the section on the d-dimensional
electrovacuum case hold also here.

\section{Boundary value problems}
\label{sectboundary}
We have shown above that if $g_{tt}$ and $\phi$ are 
functionally related, then every level-surface of $g_{tt}$ is also a 
level-surface of $\phi$. In the four-dimensional case, 
Majumdar also proved the proposition that if there exist a surface $S$
on which $g_{tt}$ and $\phi$ are both constant, and if one of the two domains
into which such a surface divides the entire spacetime is free from matter,
then, in this domain, every level-surface of  $g_{tt}$ will also be a
level-surface of $\phi$, and therefore $g_{tt}$ and $\phi$ will be
functionally related by an equation like (\ref{weylcondition}).
Here we prove that this theorem holds also in higher dimensions.
For we write Eq. (\ref{max2}) in the absence of electric charge
\beq
\partial_i\left(\frac{\sqrt{-g\,}}{V}h^{ij}\partial_j \phi\right)
=  0 \, , \label{max2free}
\eeq
where $g= -V\,h$ is the determinant of the metric.  Moreover, Eq.
(\ref{ttein}) in the absence of matter can be cast into the form
 \beq
\partial_i\left(\frac{\sqrt{-g\,}}{V}h^{ij}\partial_j V\right) = 
4\frac{{\rm d}-3}{{\rm d}-2}\,{\sqrt{-g\,}\over V}
h^{ij} \partial_i \phi\, \partial_j \phi \, . \label{tteinfree}
\eeq
Multiplying Eq. (\ref{max2free}) by $4\frac{{\rm d}-3}{{\rm d}-2}\phi$, 
subtracting
form (\ref{tteinfree}) and rearranging, we have  
\beq
\partial_i\left[\frac{\sqrt{-g\,}}{V}h^{ij}\partial_j\left( V -
2\frac{{\rm d}-3}{{\rm d}-2}\phi^2 \right)\right] = 0 \, , \label{combined}
\eeq
which has the same form of Eq. (58) of Ref. \cite{maj47}.
Now we notice that an arbitrary constant can be added to the potential 
$\phi$ in Eq. (\ref{max2free}) [and/or also to the function
$V - 2\frac{{\rm d}-3}{{\rm d}-2}\phi^2$ in Eq. (\ref{tteinfree})]. Hence, 
adding 
such a constant and multiplying (\ref{max2free}) by another constant $B$
and subtracting from (\ref{combined}) we find
 \beq
\partial_i\left[\frac{\sqrt{-g\,}}{V}h^{ij}\partial_j\left( V -
2\frac{{\rm d}-3}{{\rm d}-2}\phi^2 -B\phi -A\right)\right] = 0 \, , 
\label{combined2}
\eeq
where $A$ is a constant.
The proof of the theorem is then completed following exactly the
same steps as done by Majumdar. In particular, it follows that
$V$ and $\phi$ are functionally related by $V -
2\frac{{\rm d}-3}{{\rm d}-2}\phi^2 -B\phi -A =0$.

\section{Conclusions}
\label{conclusions}

We have investigated here the main properties of charged fluid
distributions in higher dimensional Einstein-Maxwell gravity, imposing
Weyl type and Majumdar-Papapetrou type conditions.  Several properties
of such solutions in four-dimensional spacetime are shown to hold also
in d-dimensional spacetimes. At any dimension higher than three, a
distribution of charged dust with constant matter to charge densities
ratio can be stable, independently of the shape of the body. We also
showed here that for ${\rm d}>4$ the $({\rm d}-1)$-space is conformal
to a Ricci flat space (which can be non-flat). In ${\rm d}=4$, every
solution satisfying Majumdar-Papapetrou conditions have a three
dimensional spatial sub-space conformal to a flat space, because the
three-dimensional Ricci tensor is proportional to the Riemann tensor.
There exist a vast literature on particular of Majumdar-Papapetrou
type solutions for charged dust in four-dimensional general relativity
\cite{ivanov02a}, and also including dilaton or other scalar fields
\cite{azumakoikawa}. The study of analogous problems in d-dimensional
spacetimes is of course of interest.

\section*{Acknowledgments}
This work was partially funded by Funda\c c\~ao para a Ci\^encia e
Tecnologia (FCT) through project PDCT/FP/50202/2003. 
JPSL and VTZ thank Observat\'orio Nacional-Rio de Janeiro for hospitality.

\appendix*
\section{functional relation ${\bm{p=p(\phi)}}$}

\noindent Eq. (\ref{conserveq2}) is of the form 
\beq
{\partial p\over \partial x^i} = f(x^j) {\partial \phi\over \partial x^i}\, ,
\label{conserveq3}
\eeq
where $x^i$, $i = 1, 2, ...,{\rm d}-1,$ are spacelike
coordinates in the (d$-1$)-dimensional space, and
$f(x^j) \equiv\displaystyle{\left(\rho_{\rm m}+p\right){ V'\over 2{V\,}} -
{\rho_{\rm e}\over\sqrt{V}}}\, $ is a function of the coordinates.
Multiplying both sides of Eq. (\ref{conserveq3}) by $d x^i$
and adding over $i$ results in $\;$
${({\partial p/\partial x^i})\,dx^i = 
f(x^j)\, ({\partial \phi/ \partial x^i})\,dx^i}\, ,$
which is equivalent to
\beq
dp = f(x^i) d\phi \, . \label{dphi0}
\eeq
The potential $\phi$ is also a function of the coordinates, $\phi=
\phi(x^i)$.  Assuming that $\phi$ is an invertible function of (at
least) one of the coordinates, $x^1$, say, then we can write $x^1$ in
terms of $\phi$ and of the other coordinates $x^a$, $a=2,3, ..., {\rm
d -}1$. The pressure $p$ is then a function of $\phi$ and of the remaining
spacelike coondinates $x^a$, $p = p(\phi, x^a)$. Therefore, it follows
\begin{equation}
dp = \left({\partial p\over\partial\phi}\right)_{\!x^a}d\phi + 
\left({\partial p \over\partial x^a}\right)_{\!\phi} dx^a \, , 
\label{dphi1}
\end{equation}
where $\displaystyle{\left({\partial p\over\partial \phi}\right)_{\! x^a}}$
means the derivative is done with $x^a$ held
contant, and $\displaystyle{\left({\partial p\over\partial x^a}
\right)_{\!\phi}}$ means the derivative is done with $\phi$ held constant.
Comparing the last equation to Eq. (\ref{dphi0}) it follows
\begin{eqnarray}
&&\left(\partial p \over\partial \phi\right)_{\! x^a}= f(\phi,x^a)\, ,
\label{dpdx1}\\
& & \left({\partial p \over\partial x^a}\right)_{\!\phi}
= 0\, . \label{dpdxa}
\end{eqnarray}
 Eq. (\ref{dpdx1}) gives the dependence of $p$ upon $\phi$,
while Eq. (\ref{dpdxa}) establishes that, for constant $\phi$,
the pressure $p$ does not depend on anyone of the coordinates $x^a$,
$a=2, 3, ..., {\rm d-}1$.
Finally, by repeating the above procedure 
with $x^1$ replaced, e.g., by $x^2$, it is also shown that
$\displaystyle{\left({\partial p \over\partial x^1}\right)_{\!\phi}} =0$.
So, the partial derivatives of $p$ with respect to all
the coordinates $x^i$, with $\phi$ held constant, are zero,
$$
\left({\partial p \over\partial x^i}\right)_{\!\phi} =0\,,\quad
i=1, 2, ..., {\rm  d-}1\, , $$
completing the proof.

\end{document}